\documentclass[conference]{IEEEtran}
\IEEEoverridecommandlockouts
\usepackage[comma, sort&compress,square,numbers]{natbib}
\usepackage{url}
\usepackage{hyperref}
\usepackage{amsmath,amssymb,amsfonts}
\usepackage{algorithmic}
\usepackage{graphicx}
\usepackage{textcomp}
\usepackage{xcolor}
\usepackage{authblk}
\usepackage{subcaption}
\usepackage{multirow}
\usepackage[font=small]{caption}
\usepackage{booktabs,soul}
\usepackage{flushend}

\def\BibTeX{{\rm B\kern-.05em{\sc i\kern-.025em b}\kern-.08em
    T\kern-.1667em\lower.7ex\hbox{E}\kern-.125emX}}
\begin{document}

\title{Immersive Video Compression using Implicit Neural Representations
\thanks{This work was funded by UK EPSRC (iCASE Awards), BT and the UKRI MyWorld Strength in Places Programme. High performance computational facilities were provided by the Advanced Computing Research Centre, University of Bristol.
}
}

\author[1]{Ho Man Kwan}
\author[1]{Fan Zhang}
\author[2]{Andrew Gower}
\author[1]{David Bull}
\affil[1]{\textit{Visual Information Laboratory, University of Bristol, Bristol, BS1 5DD, United Kingdom}}
\affil[1]{\textit {\{hm.kwan, fan.zhang, dave.bull\}@bristol.ac.uk}}
\affil[2]{\textit{Immersive Content \& Comms Research, BT, UK}}
\affil[2]{\textit {andrew.p.gower@bt.com}}

\maketitle

\begin{abstract}

Recent work on implicit neural representations (INRs) has evidenced their potential for efficiently representing and encoding conventional video content. In this paper we, for the first time, extend their application to immersive (multi-view) videos, by proposing MV-HiNeRV, a new INR-based immersive video codec. MV-HiNeRV is an enhanced version of a state-of-the-art INR-based video codec, HiNeRV, which was developed for single-view video compression. We have modified the model to learn a different group of feature grids for each view, and share the learnt network parameters among all views. This enables the model to effectively exploit the spatio-temporal and the inter-view redundancy that exists within multi-view videos. The proposed codec was used to compress multi-view texture and depth video sequences in the MPEG Immersive Video (MIV) Common Test Conditions, and tested against the MIV Test model (TMIV) that uses the VVenC video codec. The results demonstrate the superior performance of MV-HiNeRV, with significant coding gains (up to 72.33\%) over TMIV. The implementation of MV-HiNeRV is published for further development and evaluation\footnote{\url{https://hmkx.github.io/mv-hinerv/}}.

\end{abstract}

\begin{IEEEkeywords}
Video Compression, Immersive video, Multi-view video, Implicit neural representation, MV-HiNeRV
\end{IEEEkeywords}

\section{Introduction}
\label{sec:intro}
As part of an extended video parameter space, including higher spatial resolution, greater dynamic range, higher frame rate and wider colour gamut, new video formats have emerged to enable more immersive viewing experience with three or six degrees of freedom (3DoF or 6DoF). These underpin the development of virtual reality (VR), augmented reality (AR) or mixed reality (MR) systems \cite{bull2021intelligent}. Raw immersive video data is typically generated based on computer-generated imagery models or captured using multiple-camera systems and dedicated geometry sensors. It is then converted into different data formats, such as point clouds, multi-view texture+depth or equirectangular video, which are compressed using different video/data codecs for transmission or storage. To present or display immersive video content, the compressed data is decoded and synthesised/rendered to enable 3DoF/6DoF viewing capabilities on VR, AR or MR devices.

To standardise immersive video production and streaming, the Moving Picture Experts Group (MPEG) has developed various solutions, which support current immersive video formats including multi-view video with depth and point cloud (PC). The former is commonly used in video-based production workflows (e.g. 3D film) and the latter is mainly employed for 3D graphics-based production (e.g. 3D games). For both MVD and PC, two standards have been recently developed in the framework of MPEG-I, referred to as  MIV (MPEG Immersive Video) \cite{boyce2021miv} and V-PCC (Video-based Point Cloud Compression) \cite{graziosi2020overview}. In this paper, we focus solely on video compression for MVD. 

In the MIV pipeline, the input comprises multiple source views for both texture and depth data, together with associated source camera parameters. Both texture and depth frames are first processed to identify basic and additional views. The redundancies in the additional views are then removed, while non-redundant information is combined with the basic video and compressed using a standard video encoder (e.g., HEVC HM \cite{sullivan2012overview}  or VVC VVenC \cite{vvenc}). It is noted that, while it is convenient to employ a conventional video codec for texture and depth video compression, such codecs are not designed to exploit the redundancy within multiple views, and MIV achieve this by extracting basic views and removing redundant information in additional views. 

In the realm of learning-based video compression, deep neural networks have been deployed to enhance the traditional coding pipeline \cite{song2017neural,afonso2018video,zhang2021video,ma2020mfrnet,zhang2021vistra2} or build a fully learnable framework to achieve end-to-end optimisation \cite{lu2019dvc, li2021deep, agustsson2020scale,li2023neural}. More recently,  Implicit Neural Representations (INRs)  \cite{chen2021nerv, chen2023hnerv, lee2023ffnerv, kwan2023hinerv, gomes2023entropy} have shown great potential to achieve comparable compression performance to both standard and other learning-based video codecs, importantly also achieving relatively fast decoding speeds. An INR is a neural network that is optimised to map coordinates to pixels during video encoding, reconstructing the entire video at the decoder by performing inference. However, existing INR-based codecs were designed for encoding single-view videos; the application to multi-view video coding has not previously been investigated.

To this end, we investigate the use of INRs for immersive video compression, and propose a new INR-based multi-view video codec, MV-HiNeRV. The proposed approach is an extension of a state-of-the-art INR-based video codec, HiNeRV \cite{kwan2023hinerv}, which was developed for conventional video compression. Specifically, for multi-view video coding, MV-HiNeRV learns a unique set of feature grids for each view, with shared network parameters across views. The sharing of network parameters effectively exploits the spatio-temporal redundancy among different views.

To evaluate its coding efficiency,  MV-HiNeRV has been tested on MIV CTC test sequences \cite{mivctc} and compared against the Test Model of MIV (TMIV) \cite{tmiv}. The results show its significant performance improvement over the original TMIV, with an average bit rate saving of 46.92\%. To the best of our knowledge, MV-HiNeRV is the first INR-based immersive video codec, and the results demonstrate the promise of using INR-based models for immersive video compression.

\section{Method}
Implicit Neural Representation (INR)-based video compression has been shown to outperform or compete with various advanced standard codecs in terms of coding efficiency \cite{chen2021nerv, chen2023hnerv, lee2023ffnerv, he2023towards, kwan2023hinerv, gomes2023entropy}. In the related works, a single instance of INR has shown capability of encoding a large number of video frames, for example, up-to 600 frames for the sequences in the UVG dataset\cite{mercat2020uvg}. Due to its ability to exploit redundancies between frames, this capability is likely to benefit the compression of immersive videos which exhibit additional redundancies through multiple views. Based on this observation, we propose an INR-based codec, MV-HiNeRV (a multi-view version of HiNeRV \cite{kwan2023hinerv}) for encoding multi-view videos with the aim of achieving improved coding performance.

\subsection{MV-HiNeRV}
\begin{figure}
    \centering
    \begin{subfigure}{.9\linewidth}
        \centerline{\includegraphics[width=1.0\linewidth]{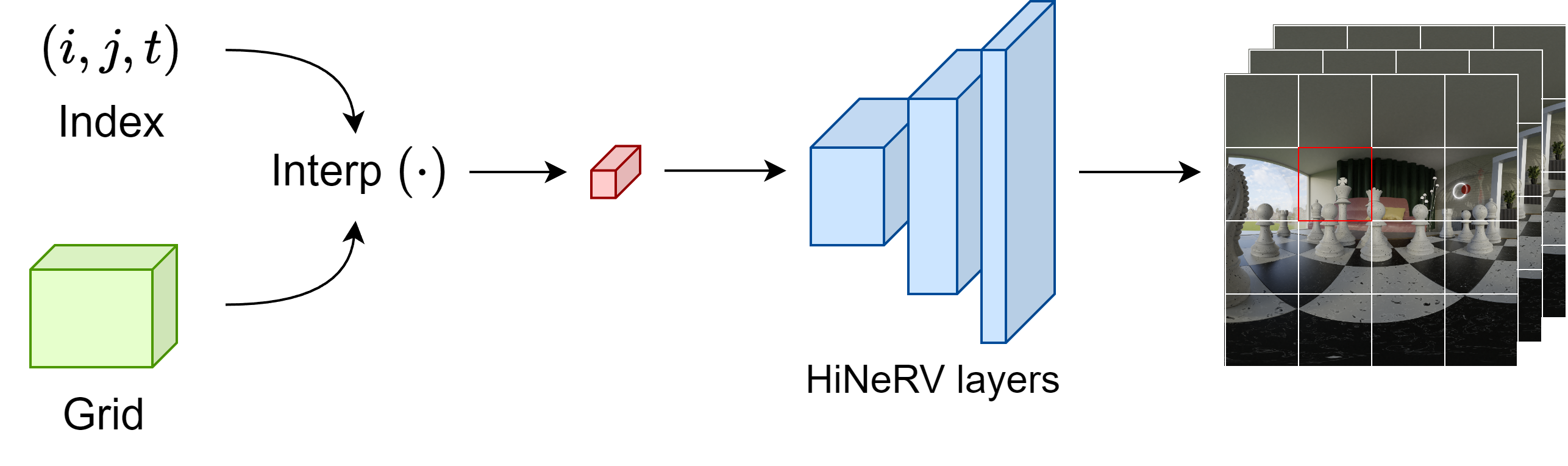}}
        \caption{}
        \label{fig:single_view}
    \end{subfigure}
    \begin{subfigure}{.9\linewidth}
        \centerline{\includegraphics[width=1.01555\linewidth]{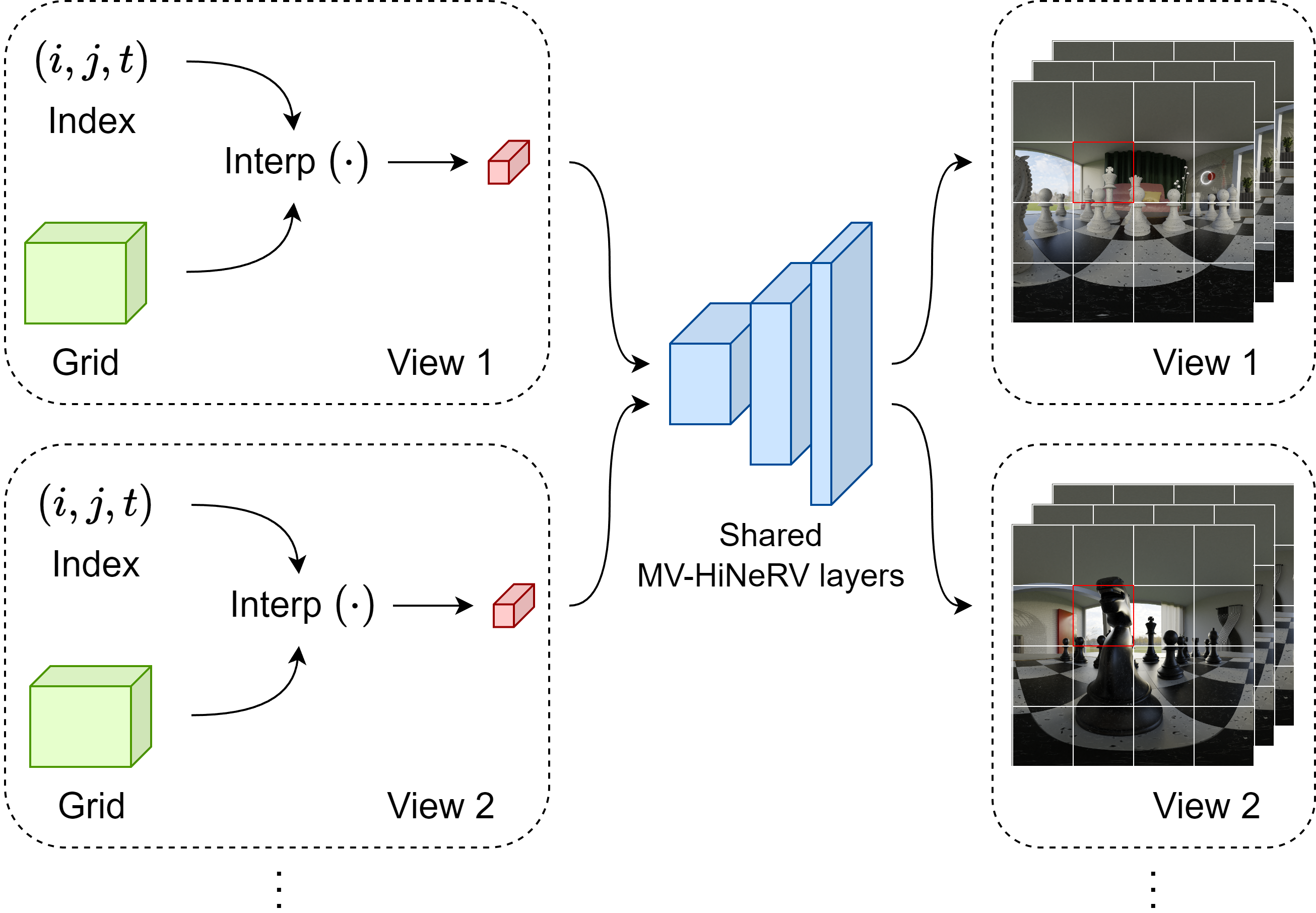}}
        \caption{}
        \label{fig:multi_view}
    \end{subfigure}
    \caption{(a). \textbf{HiNeRV.} In HiNeRV, an input patch can be obtained by interpolation from the feature grids, and the network layers output the corresponding video patch. (b). \textbf{The proposed MV-HiNeRV.} In MV-HiNeRV, each view is represented by a dedicated set of feature grids, and the network layers are shared across views. This allow efficient multi-view video coding as it exploiting the spatial, temporal and view redundancy simultaneously.}
    \vspace{-10pt}
\end{figure}

Existing multi-view video coding methods either compress multiple views by leveraging inter-view redundancy \cite{vetro2011overview, tech2016overview}, or apply view pruning \cite{boyce2021miv}. As mentioned above, by utilising INR models, it is possible to directly compress a large number of video views with a high coding efficiency. Moreover, using INR models for exploiting redundancy is potentially time efficient, since the encoding process is the model training, which its length has a sub-linear relationship with the number of input frames or videos; this differs from conventional and other learning-based coding methods.

In this work, to adapt HiNeRV for multi-view video compression, we modified the original INR model by storing a separated set of feature grids for each view, where the network parameters, e.g., those for the convolutional and fully connected layers, are shared among views, in addition to the spatial and temporal dimensions. This allows simultaneous exploitation of redundancy in all three dimensions. Fig. \ref{fig:single_view} and \ref{fig:multi_view} show the difference between the original HiNeRV and the proposed multi-view variant.

Following the approach in  \cite{kwan2023hinerv}, assuming that we encode a multi-view video with $K$ views, we denote the $k$-th view as $V_{k}$, such that $0 \leq k < K$. For simplicity, we assume that the video size is the same for $V_0$...$V_{K-1}$, i.e., $T \times H \times W \times C$, where $T$, $H$, $W$ and $C$ are the temporal resolution, spatial resolution and the number of channels, respectively.

With a patch size $M\times M$, MV-HiNeRV represents each video in multiple patches, where each patch is denoted by its  patch index $(i, j, t)$, where $0 \leq t < T$, $0 \leq j<\frac{H}{M}$ and $0\leq i < \frac{W}{M}$.
To compute the output of the $k$-th view, MV-HiNeRV first computes the input feature map, i.e., the encoding obtained from interpolation with the multi-temporal resolution grids \cite{lee2023ffnerv} from the $k$-th view, $\gamma_{k, base}$; it then applies a stem convolutional layer $F_{stem}$, to obtain the first stage feature map:
\begin{equation}
    X_{0} = F_{stem}(\gamma_{k, base}(i, j, t)).
\end{equation}
Subsequently, $N$ HiNeRV blocks are progressively processed using the feature maps for both upscaling and transformation, where we denote the output of the $n$-th HiNeRV block as $X_n$, $0 < n \leq N$:
\begin{equation}
    \label{eq:X_n}
    \begin{split}
        X_n = & F_n(U_n(X_{n - 1}) + F_{enc}(\gamma_{k, n}(i, j, t))),  0 < n \leq N
    \end{split}
\end{equation}
It should be noted that the hierarchical encoding \cite{kwan2023hinerv} in MV-HiNeRV is also view dependent, as the grids are compact and only need a small number of parameters. The hierarchical encoding at the $k$-th view of the $n$-th block is denoted as $\gamma_{k, n}$. In MV-HiNeRV, we simply follow HiNeRV, where we use the ConvNeXt \cite{liu2022convnet} block as the internal layer in the HiNeRV blocks.

Finally, a linear head layer $F_{head}$ is applied to transform the feature maps $X_{N}$ into the target output space:
\begin{equation}
    Y = F_{head}(X_N),
\end{equation}
Unlike HiNeRV, MV-HiNeRV  encodes both the texture and depth maps at the same time, hence we produce a four channel (RGB and D) output with the linear head layer.

During model training, we randomly sample video patches from all views and frames \cite{kwan2023hinerv}. We also perform training with overlapped patches in the feature space; this supports training with patches while enhancing the encoding quality.

\subsection{Weight Quantisation and Entropy Regularisation}
\textbf{Weight quantisation.} Following \cite{gomes2023entropy}, we applied the learned quantisation and entropy regularisation to reduce the model size. Specifically, for a parameter vector $\boldsymbol{\theta} = \{\theta_i\}$ with a trainable quantisation step vector $\boldsymbol{\delta} = \{\delta_i\}$, the quantised parameter $\boldsymbol{\hat{\theta}}  = \{\hat{\theta_i}\}$ is computed as follows:
\begin{equation}
    \hat{\theta_i} = \delta \times \lfloor \frac{\theta_i}{\delta_i} \rceil.
\end{equation}
We follow the model compression implementation in \cite{tfc_github}, where the logarithm of $\delta$ is learned instead of $\delta$. The actual step sizes are also shared between a subset of the parameters from the same layer, in order to reduce the overhead. For instance, the same step size is employed for a row or a column in the weight of the linear layer. Since quantisation is a non-differentiable operation, we use Quant-Noise \cite{fan2020training} as in HiNeRV \cite{kwan2023hinerv} in the training and employ the quantisation during evaluation.

\textbf{Entropy regularisation.} To train the model parameters and the quantisation step sizes at the same time, we follow the common practice in lossy compression where  entropy regularisation is used to optimise both the rate and distortion jointly, incorporating a Lagrangian multiplier, $\lambda$:
\begin{equation}
L = R + \lambda D,
\end{equation}
Here, the rate term $R$ is defined as the sum of the negative log likelihood of the quantised weights, i.e., the lower bound of the total code length, and the distortion term $D$ can be calculated using different loss functions such as the MSE loss. For a model with a quantised parameter vector $\boldsymbol{\hat{\theta}}  = \{\hat{\theta_i}\}$, we compute the rate term by
\begin{equation}
    R = \sum_{i=0}^{|\boldsymbol{\hat{\theta}}|} -\mathrm{log}_{2} p(\hat{\theta_i})
\end{equation}

In particular, we use a multivariate Gaussian \cite{balle2018variational, minnen2018joint} for estimating the weight distribution and thus the rate term. Unlike other learning-based video codecs \cite{lu2019dvc, li2021deep}, the distribution of the symbols, can be computed easily because the symbols are simply the quantised network parameters. Thus, we compute the Gaussian parameters directly in each training step, and encode them along with with the model parameter bitstream. To estimate the rate of each symbol, we also use the continuous approximation of the quantised variables in \cite{balle2017end}. We only apply regularisation to the feature grids and the weights of the linear and convolutional layers.

Previous work \cite{gomes2023entropy} employed a non-parametric model \cite{balle2018variational} for estimating the distribution, introducing additional parameters that require training. This may increase complexity, so in this work we simply assume Gaussian distribution which is effective in practice.

\subsection{Model Training Pipeline}
Regarding model training, previous work in \cite{gomes2023entropy} adopted a two stage approach, in which two models are trained without regularisation in Stage 1, while multiple models are trained in Stage 2 with regularisation for different rate points, achieved by adjusting the weight between the distortion and rate terms. However, we found that the above approach will lead to sub-optimal rate-distortion performance when the weight of the rate term is heavier, i.e., for obtaining the more compact models. Hence, in this paper, we modified this two-stage approach \cite{gomes2023entropy}, but using a different value for $\lambda$, and a different set of the model hyper-parameters for each target rate point model, as in some other INR-based works \cite{chen2021nerv, chen2023hnerv, lee2023ffnerv, he2023towards, kwan2023hinerv}. In Stage 1, we train one MV-HiNeRV model without any regularisation which allows faster convergence. In Stage 2 we manually initialise the step sizes to a value such that the quantisation width to 7 at the beginning. Then, we train the model with entropy regularisation; where we found that using a smaller learning rate for network parameters and a larger one for the step size is actually more effective for preserving the model quality while reducing the size. We also apply a linear scheduling for the quantisation noise rate \cite{fan2020training}.  In our approach, only 300 epochs are used for Stage 1 training without regularisation, and 60 epochs for Stage 2, while 1200 and 300 epochs are used in \cite{gomes2023entropy}, respectively. Most of the training configurations for MV-HiNeRV are the same as those in \cite{kwan2023hinerv}.

\section{Experiments}
We employed six mandatory test sequences in the MIV Common Test Conditions (CTC) \cite{mivctc} for testing. We compared MV-HiNeRV with the MPEG Test Model (TMIV) \cite{tmiv} using the main anchor configuration. TMIV employs VVenC \cite{vvenc} in Random Access mode to encode both texture and depth information. Following the CTC, we performed experiments with the specified start frames and encoded 65 frames in each sequence. For MV-HiNeRV, we converted the original YUV 4:2:0 texture frames into the RGB colour space, then concatenated them with the depth frames as the input.

After decoding using MV-HiNeRV, we convert the output of MV-HiNeRV back to the original YUV 4:2:0 format, and apply the same view synthesizer in TMIV to generate both source and pose trace views for evaluation, as described in the MIV CTC \cite{mivctc}. We use both PSNR and IV-PSNR \cite{dziembowskiMSG2022ivpsnr} to evaluate video quality, and employ range coding for entropy coding to obtain the actual rate. Metrics are computed with regard to the best reference \cite{mivctc}.

For MV-HiNeRV training, we use MSE loss in the RGBD space as the distortion term $D$. 
It should be noted that the reconstruction quality is highly dependent on the accuracy of the depth map. Computing MSE loss in the depth channel may result in insufficient precision due to the varying range of depth values across different scenarios. As a result, we normalise the depth map of each sequence to [0, 1]. Detailed network and training configurations are provided in the code repository.

\subsection{Quantitative and Qualitative Results}

\begin{figure}[t]
    \centering
    \scriptsize
    \begin{subfigure}[t]{0.43\linewidth}
        \centering
        \includegraphics[width=\linewidth]{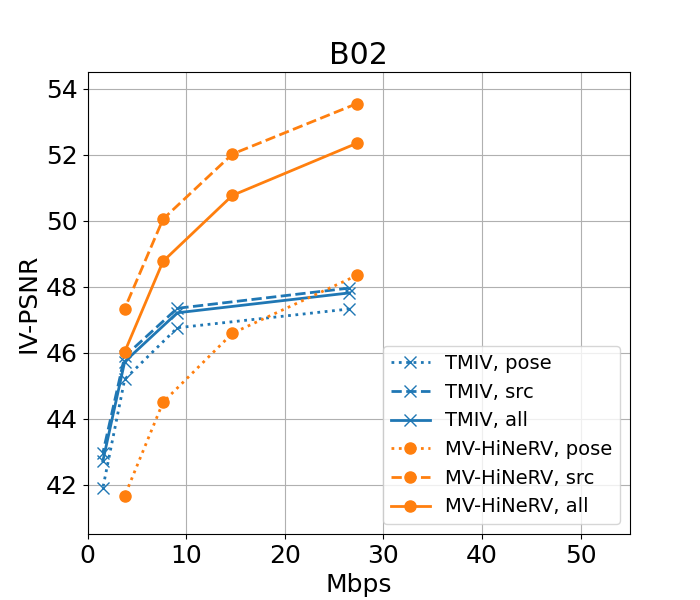}
     \end{subfigure}
    \begin{subfigure}[t]{0.43\linewidth}
        \centering
        \includegraphics[width=\linewidth]{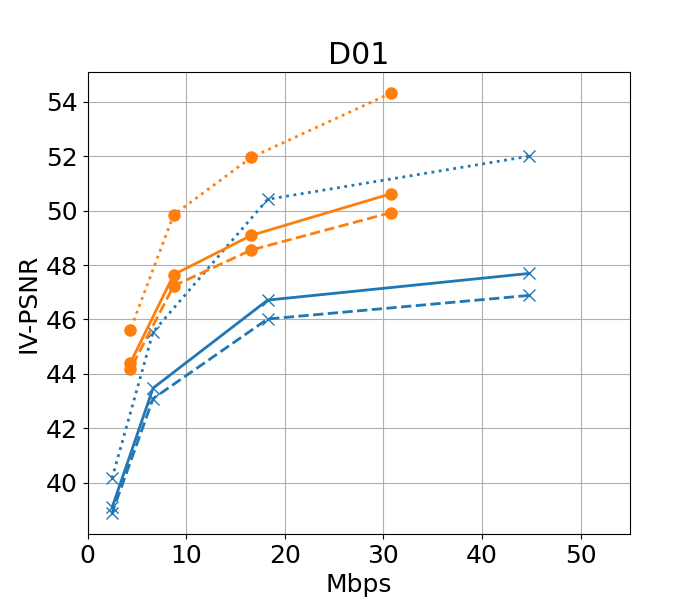}
     \end{subfigure}
    \\
    \begin{subfigure}[t]{0.43\linewidth}
        \centering
        \includegraphics[width=\linewidth]{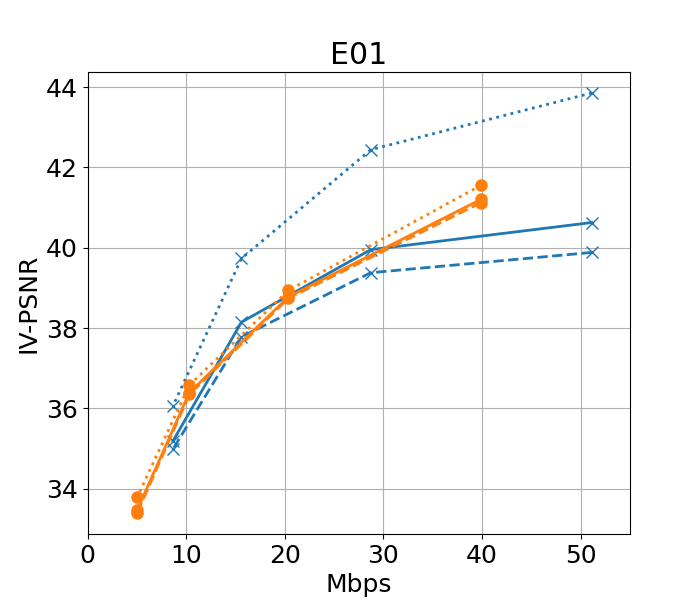}
     \end{subfigure}
    \begin{subfigure}[t]{0.43\linewidth}
        \centering
        \includegraphics[width=\linewidth]{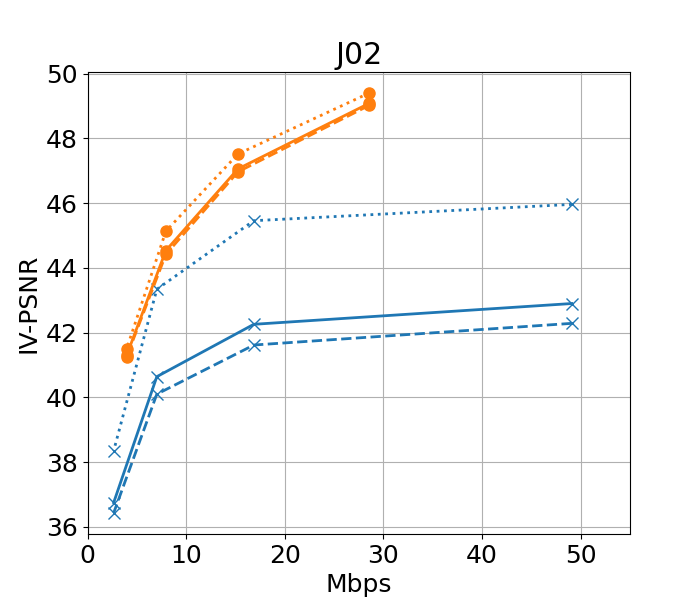}
     \end{subfigure}
    \\
    \begin{subfigure}[t]{0.43\linewidth}
        \centering
        \includegraphics[width=\linewidth]{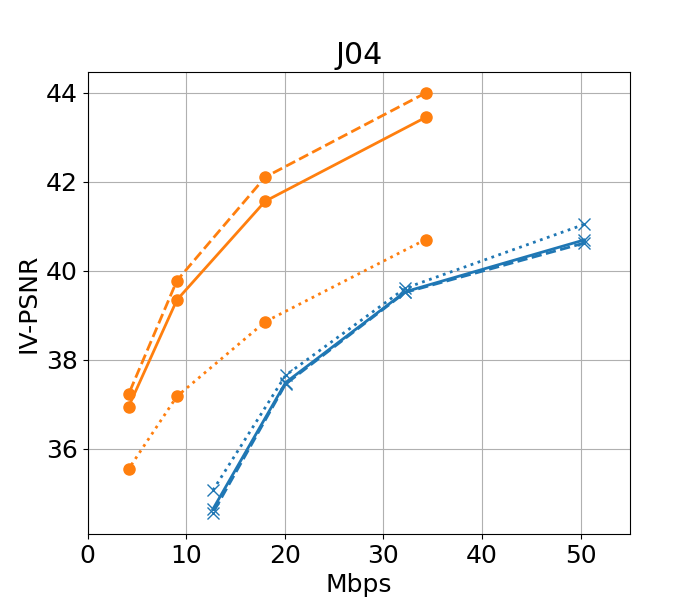}
     \end{subfigure}
    \begin{subfigure}[t]{0.43\linewidth}
        \centering
        \includegraphics[width=\linewidth]{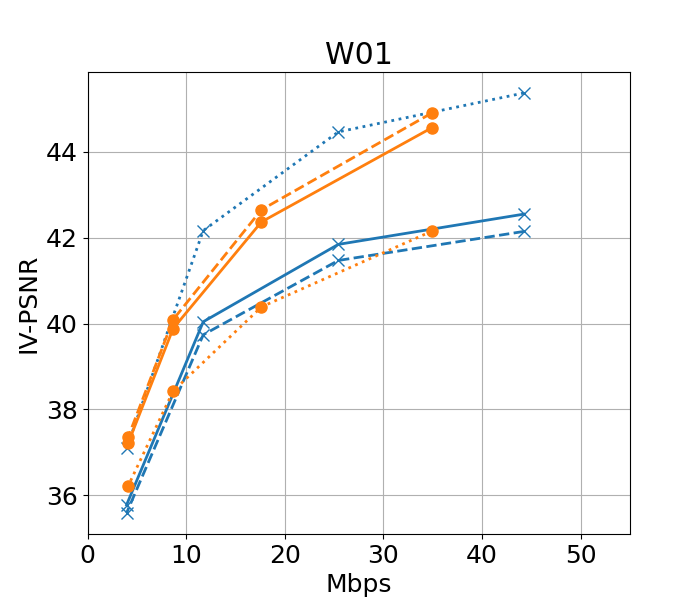}
     \end{subfigure}
    \caption{IV-PSNR results of TMIV and MV-HiNeRV on the MIV CTC test sequences. \textbf{\textit{pose}:} Pose trace views. \textbf{\textit{src}}: Source views. \textbf{\textit{all}}: The average of pose trace/source views.}
    \label{fig:results}
    \vspace{-10pt}
\end{figure}

TABLE \ref{tab:bd_rate} summarises compression results for MV-HiNeRV compared to TMIV in terms of BD-rate values measured in PSNR/IV-PSNR \cite{dziembowskiMSG2022ivpsnr} (the average among both reconstructed source and pose trace views).  MV-HiNeRV performs significantly better than TMIV on all six test sequences, achieving up to 72.33\% bit rate savings (measured by Bj{\o}ntegaard Delta Rate, BD-rate \cite{BD} using IV-PSNR). This performance is also confirmed by Fig. \ref{fig:results}, which plots the rate-PSNR curves for all sequences (including both the reconstructed views at the source and the pose trace views).

We also measured the bit rate distribution, and we found that the shared parameters accounted for more than 60\% of the total rate on average. Given the high quantitative performance, this suggest that in MV-HiNeRV, the shared parameters are an effective shared representation between views.

\begin{table}[t]
\caption{BD-Rate (\%, measured in PSNR/IV-PSNR) results of MV-HiNeRV on the MIV CTC test sequences (for both source and pose trace views). The anchor is TMIV.}
\centering
\resizebox{.9\linewidth}{!}{
    \begin{tabular}{c|c|c|c|c|c|c|c}
        \toprule
        \textbf{Metric} & \textbf{B02} & \textbf{D01} & \textbf{E01} & \textbf{J02}& \textbf{J04} & \textbf{W01} & \textbf{Overall} \\
        \midrule
        PSNR & -17.60 & -65.93 & -36.59 & -59.96 & -80.03 & -35.48 & -49.27 \\
        \midrule
        IV-PSNR & -38.11 & -61.08 & -6.28 & -70.21 & -72.33 & -33.50 & -46.92 \\
        \bottomrule
    \end{tabular}
}
\label{tab:bd_rate}
\vspace{-10pt}
\end{table}

\begin{figure}[btbp]
     \centering
     \scriptsize
     \begin{subfigure}[t]{.95\linewidth}
        \centering
        \includegraphics[width=\linewidth]{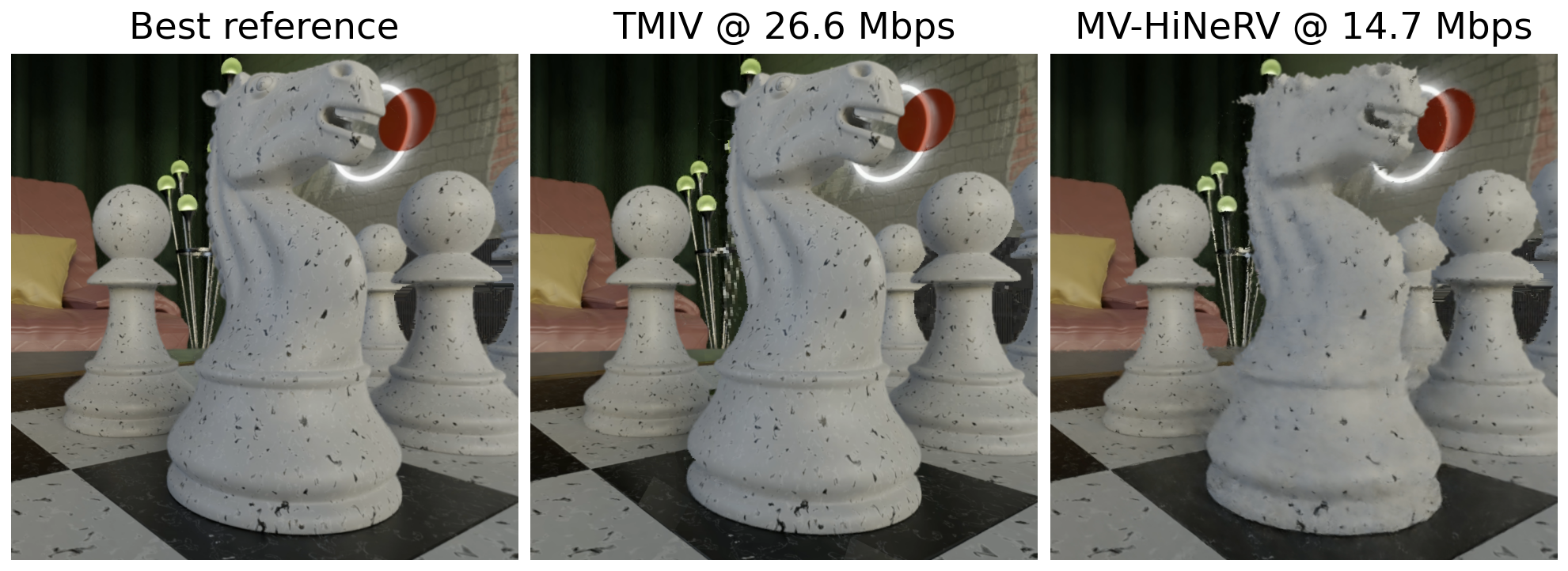}
        \vspace{-13pt}
        \caption{B02 - p03}
     \end{subfigure}
     \begin{subfigure}[t]{.95\linewidth}
        \centering
        \includegraphics[width=\linewidth]{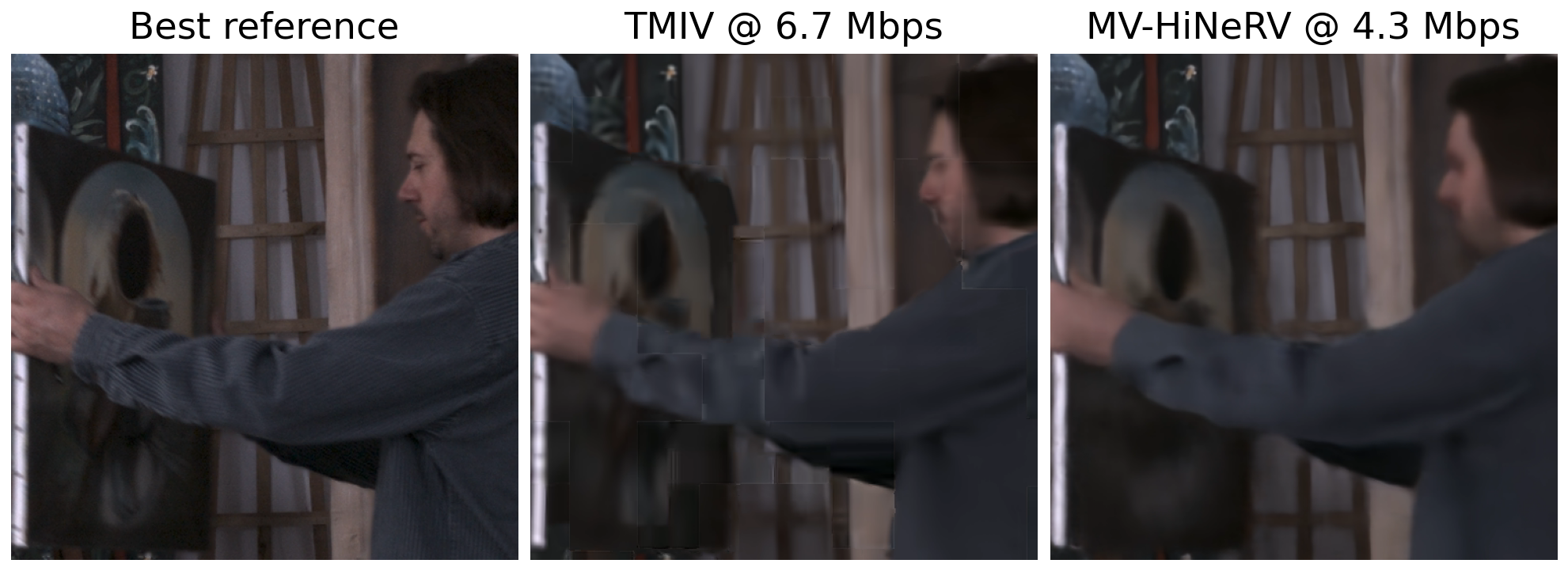}
        \vspace{-13pt}
        \caption{D01 - v13}
     \end{subfigure}
     \begin{subfigure}[t]{.95\linewidth}
        \centering
        \includegraphics[width=\linewidth]{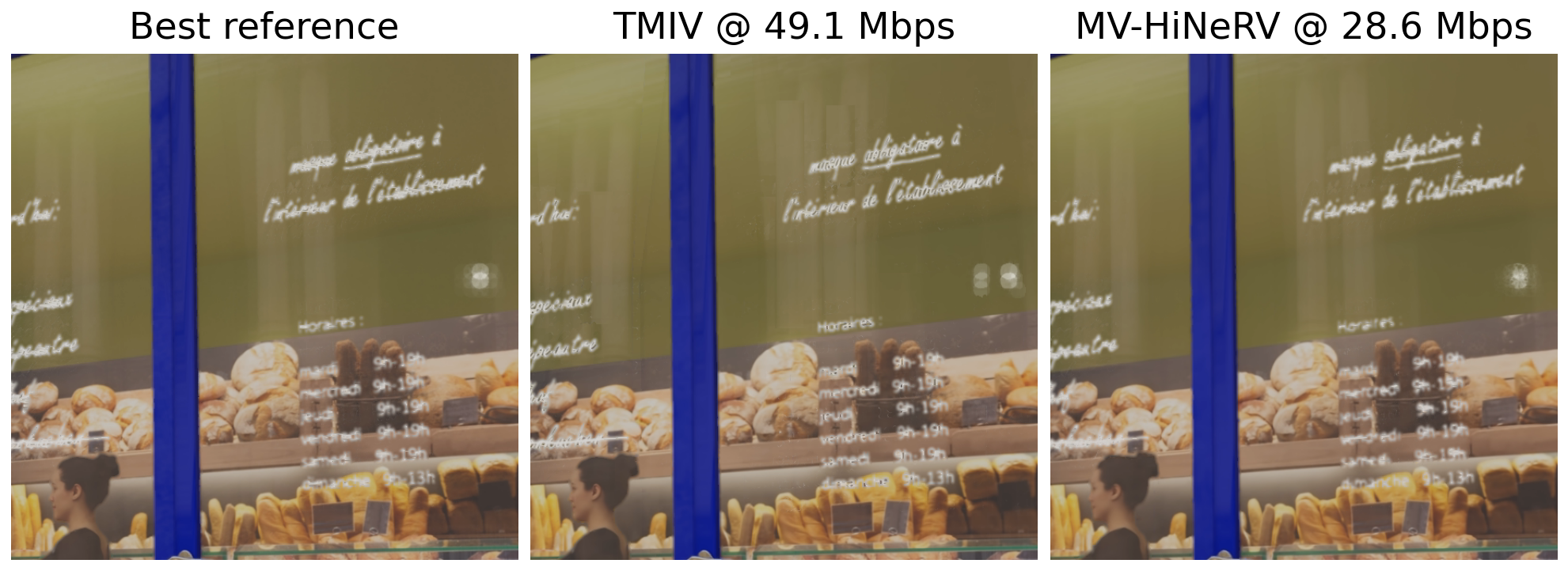}
        \vspace{-13pt}
        \caption{J02 - p02}
     \end{subfigure}
     \begin{subfigure}[t]{.95\linewidth}
        \centering
        \includegraphics[width=\linewidth]{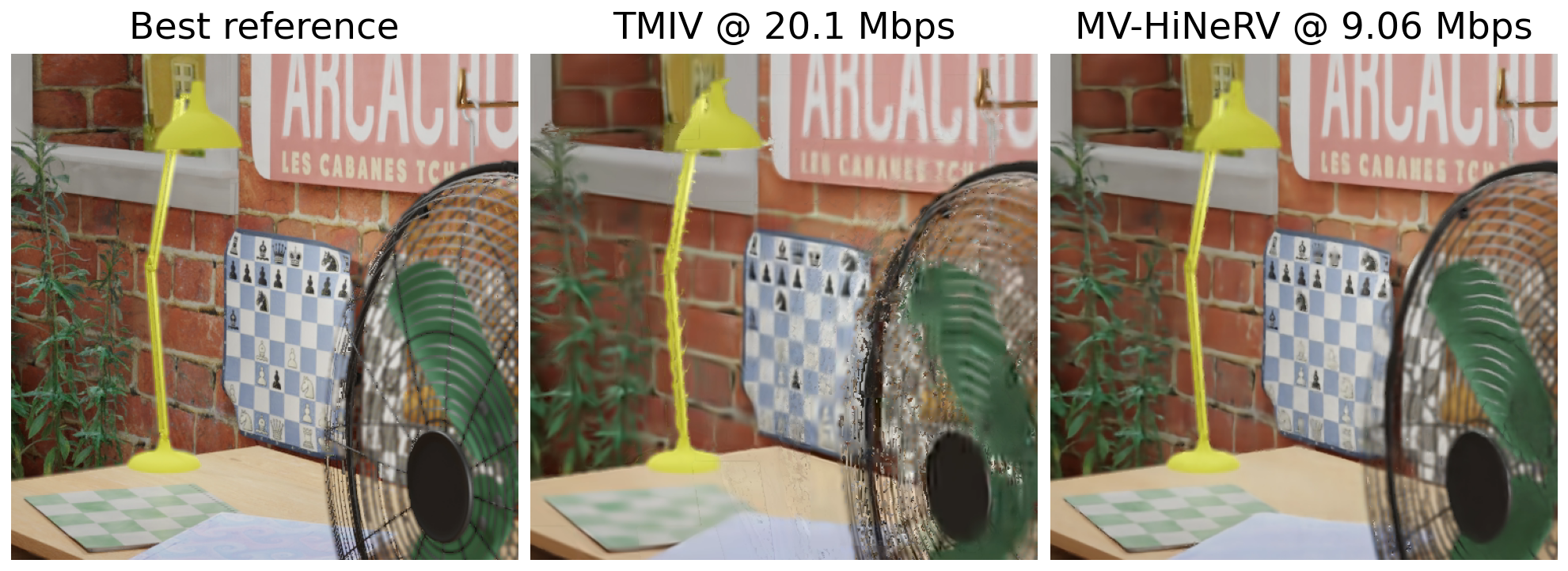}
        \vspace{-13pt}
        \caption{J04 - v07}
     \end{subfigure}
     \\
     \caption{Example of synthesised views (cropped) of the CTC test sequence \cite{mivctc}. \textbf{\textit{pXX}}: pose trace views. \textbf{\textit{vXX}}: source views. }
    \label{fig:samples}
    \vspace{-10pt}
\end{figure}

For perceptual quality assessment, we provide visual examples of the synthesised views from MIV and MV-HiNeRV (using the same viewer synthesiser) in Fig. \ref{fig:samples}. It can be observed that, in general, the MV-HiNeRV produces high-quality frames with rich spatial details. More importantly, the outputs contain fewer visual artefacts as in TMIV, which may be due to the MIV preprocessing. In these cases, the bit rates for MV-HiNeRV examples are similar or even smaller than that for MIV. The only issue we noticed is that MV-HiNeRV perform worse in cases where high depth map precision is required, such as for the edges of the foreground objects in sequence B02.

In terms of complexity, the encoding process with MV-HiNeRV is relatively slow, similar to existing INR-based methods \cite{chen2021nerv, chen2023hnerv, lee2023ffnerv, he2023towards, kwan2023hinerv, gomes2023entropy}. For example, encoding the sequence W01, which comprises 21 full HD views, each consisting of 65 frames, can take up to 26 hours on a computer with a V100 GPU. While decoding with INRs is typically fast, in our case, the entire decoding process is limited by  the TMIV view synthesizer. For the same sequence, decoding all source views with MV-HiNeRV can be completed in less than 2 minutes, but rendering a synthesised view with the TMIV view synthesizer takes approximately 12 minutes.

\section{Conclusion}
In this paper, we proposed an INR-based codec, MV-HiNeRV, for encoding multi-view videos. MV-HiNeRV is an extension of HiNeRV, which was developed for single-view video compression. The proposed approach learns a different set of feature grids for each view and shares the model parameters among all views,  effectively exploiting the inter-view redundancy. MV-HiNeRV has been evaluated against the MPEG MIV Test Model, TMIV, and achieved significant performance improvement, with 46.92\% average coding gain. The results demonstrate the significant potential of INR-based video codecs for the compression of immersive video formats. Future work should focus on better integration of viewer synthesiser with INR-based immersive video codecs.

\small
\bibliography{egbib}
\bibliographystyle{IEEEtran}

\end{document}